\documentclass[reprint,
superscriptaddress,
amsmath,amssymb,aps,showkeys,showpacs,
twoside, final,secnumarabic,
nofootinbib]{revtex4-2}

\usepackage[paperwidth=205mm,paperheight=290mm,top=17mm,bottom=25mm,
inner=17mm,outer=17mm,
twoside]{geometry}

\usepackage{cmap} 
\usepackage[T1,T2A]{fontenc}
\usepackage[utf8]{inputenc}
\usepackage[russian,english]{babel}
\usepackage{color}
\usepackage{graphicx}
\usepackage{dcolumn}
\usepackage{bm} 
\usepackage[unicode=true,colorlinks=true,linkcolor=magenta, urlcolor=blue, citecolor = blue,breaklinks]{hyperref}
\usepackage{multirow}
\usepackage{url}
\usepackage{breakurl}
\DeclareGraphicsExtensions{.eps}

\newcount\issue
\newcount\Vol
\newcount\numb
\usepackage{fancyhdr} 
\def\Vol{\textbf{78}}
\def\numb{x}
\begin{document}

\title{JOURNAL SECTION OR CONFERENCE SECTION\\[20pt]
Charmed Baryons at Belle and Belle II} 

\def\addressa{HSE, }

\author{\firstname{A.O.}~\surname{Mufazalova}}
\email[E-mail: ]{mufazalova.a.o@hse.ru }
\affiliation{National Research University Higher School of Economics, 20 Myasnitskaya Street, Moscow 101000, Russia}

\received{xx.xx.2025}
\revised{xx.xx.2025}
\accepted{xx.xx.2025}

\begin{abstract}
We present the recent results on charm baryon production and decays at Belle and Belle II experiments using a 1.4\,ab$^{-1}$ data sample of  $e^+ e^-$ collision collected at center-of-mass energies near the $\Upsilon(nS)$ resonances. The collected data contain a large number of $e^+ e^- \to c \bar{c}$ events that produce charmed baryons. We report on several studies of the $\Lambda_c$ and $\Xi_c$ decays to determine their branching fractions.
\\

\end{abstract}

\pacs{10}\par
\keywords{Charmed baryons\\[5pt]}

\maketitle
\thispagestyle{fancy}


\section{Introduction}\label{intro}

The study of charmed baryon decays offers an opportunity to explore both strong and weak interactions and their interplay. 
Decays of charmed baryons have a large contribution from non-spectator diagrams, where a charm quark interacts directly with a light quark via a $W$ boson, changing the quark flavor.

\section{\label{sec:level1} Overview of selected results}

In this paper, we present an overview of some recent results on charmed baryons from the Belle and Belle~II experiments. 
All the results have statistical and systematic uncertainty, respectively.

\subsection{\label{sec:level2} Measurements of the branching fractions of $\Xi^+_c \to \Sigma^+ K^0_S$, $\Xi^0 \pi^+$ and $\Xi^0K^+$}

\begin{figure*}
\includegraphics[width = 1\linewidth]{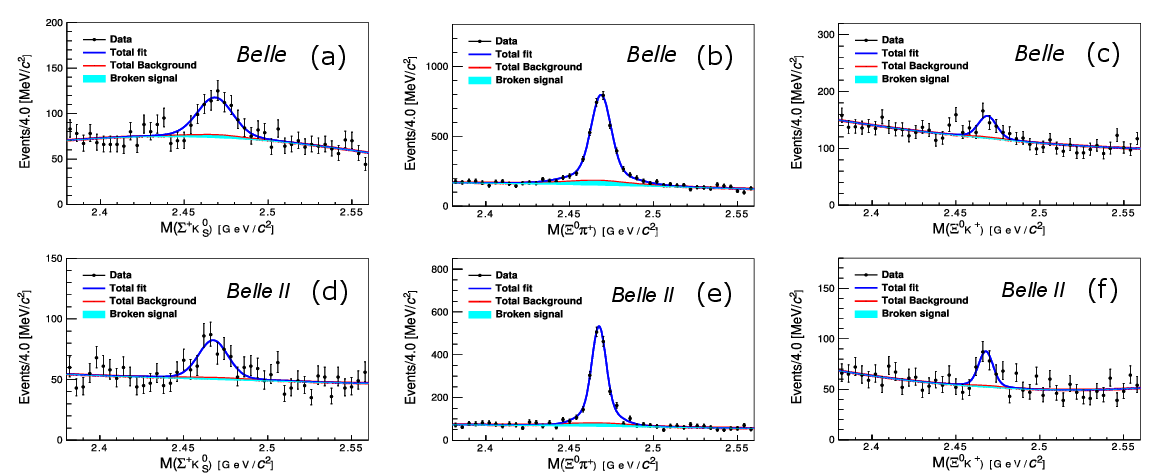} 
\caption{\label{fig:wide_Xi_c_p}
Invariant mass spectra of $\Xi_c^+$ candidates from $\Xi_c^+ \to \Sigma^+K_S^0$ (a, d) $\Xi_c^+ \to \Xi^0 \pi^+ $ (b, e) and $\Xi_c^+ \to \Xi^+K_S^+$ (c, f) decays reconstructed in (top) Belle and (bottom) Belle II data. 
}
\end{figure*}

Here we report on the first observation of the Cabibbo suppressed decays $\Xi_c^+ \to \Sigma^+ K_S^0$ and $\Xi_c^+ \to \Xi^0K^+$  and improved measurement of the Cabibbo allowed decay $\Xi_c^+ \to \Xi^0\pi^+$.
The analysis is described in detail elsewhere~\cite{JHEP_Xi_c_plus}. The corresponding combinations of strange hyperon and meson have been selected to form $\Xi^+_c$ candidates, and their invariant mass spectra are shown in Fig.~\ref{fig:wide_Xi_c_p}. 

All data spectra are peaking around the nominal $\Xi_c^+$ mass~\cite{PDG}. 
The signal yields were extracted from the fit to these spectra, and the branching fractions are calculated as follows

\begin{subequations}
\label{eq:whole}
\begin{eqnarray}
        \frac{\mathcal{B}(\Xi_c^+ \to \Sigma^+K_S^0)}{\mathcal{B}(\Xi_c^+ \to \Xi^- \pi^+ \pi^+)} = 0.067\pm0.007\pm0.003, \\
      \frac{\mathcal{B}(\Xi_c^+ \to \Xi^0\pi^+)}{\mathcal{B}(\Xi_c^+ \to \Xi^- \pi^+ \pi^+)} = 0.248\pm0.005\pm0.009,  \\
          \frac{\mathcal{B}(\Xi_c^+ \to \Xi^0 K^+)}{\mathcal{B}(\Xi_c^+ \to \Xi^- \pi^+ \pi^+)} = 0.017\pm0.003\pm0.001. 
\end{eqnarray}
\end{subequations}

Compared to theoretical predictions \cite{t6,t7,t8} for the similar decays, the value of $\mathcal{B}(\Xi_c^+ \to \Sigma^+K_S^0)$ lies lower. The $\mathcal{B}(\Xi_c^+ \to \Xi^0\pi^+)$ agrees with theoretical predictions and the previous CLEO result \cite{CLEO_2}. And the $\mathcal{B}(\Xi_c^+ \to \Xi^0 K^+)$ is below the central values predicted by the most theoretical papers.

\subsection{\label{sec:level2} Measurements of the decay $\Lambda^+_c \to pK^0_S\pi^0$}

Another result we highlight here is a precise measurement of the ratio $\mathcal{B}(\Lambda^+_c \to pK^0_S\pi^0)/\mathcal{B}(\Lambda^+_c \to pK^-\pi^+)$~\cite{PRD_Lamb}.
This study examines the isospin properties of the weak $c \to s$ transition. 

\begin{figure}[h!]
\includegraphics[width=0.8\linewidth]
{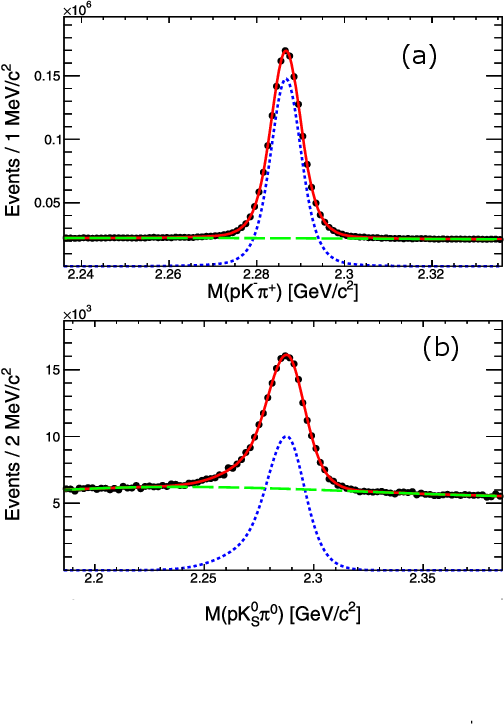}
\caption{\label{fig:Lamb} The invariant mass spectra of $\Lambda_c^+$ candidates for $\Lambda_c^+ \to pK^-\pi^+$ (a) and $\Lambda_c^+ \to pK^0_S\pi^0$ (b). The total fit is represented by a solid red curve, the signal by a dashed blue curve, and the background by a long-dashed green curve.
}
\end{figure}

The invariant mass spectra of $pK^0_S\pi^0$ and $pK^-\pi^+$ are shown in Fig.~\ref{fig:Lamb}.
Both spectra have prominent peaks at the $\Lambda_c^+$ mass. 
The fitted yields are then corrected for the efficiencies to calculate the ratio of corresponding branching fractions:

\begin{equation}
    \frac{\mathcal{B}(\Lambda_c^+ \to pK^0_S \pi^0)}{\mathcal{B}(\Lambda_c^+ \to p K^- \pi^+)} =
    0.339\pm0.002\pm 0.009. 
\end{equation}

This measurement improves the uncertainty of the previous CLEO result by a factor of five \cite{CLEO}.

\subsection{\label{sec:level2} Measurements of the decays $\Xi^0_c \to \Xi^0\pi^0$, $\Xi^0\eta$ and $\Xi^0\eta'$ and the asymmetry parameter in $\Xi^0_c \to \Xi^0\pi^0$}

Finally, we discuss the results \cite{JHEP_Xi_c_0} of the first measurements of the Cabibbo allowed decays  $\Xi_c^0 \to \Xi^0\pi^0$, $\Xi_c^0 \to \Xi^0 \eta$, $\Xi_c^0 \to \Xi^0\eta'$ and the asymmetry parameter $\alpha$ of $\Xi^0_c \to \Xi^0 \pi^0$ decay arising from interference between the parity-violating and parity-conservating amplitudes.

\begin{figure*}
\includegraphics[width = 1\linewidth]{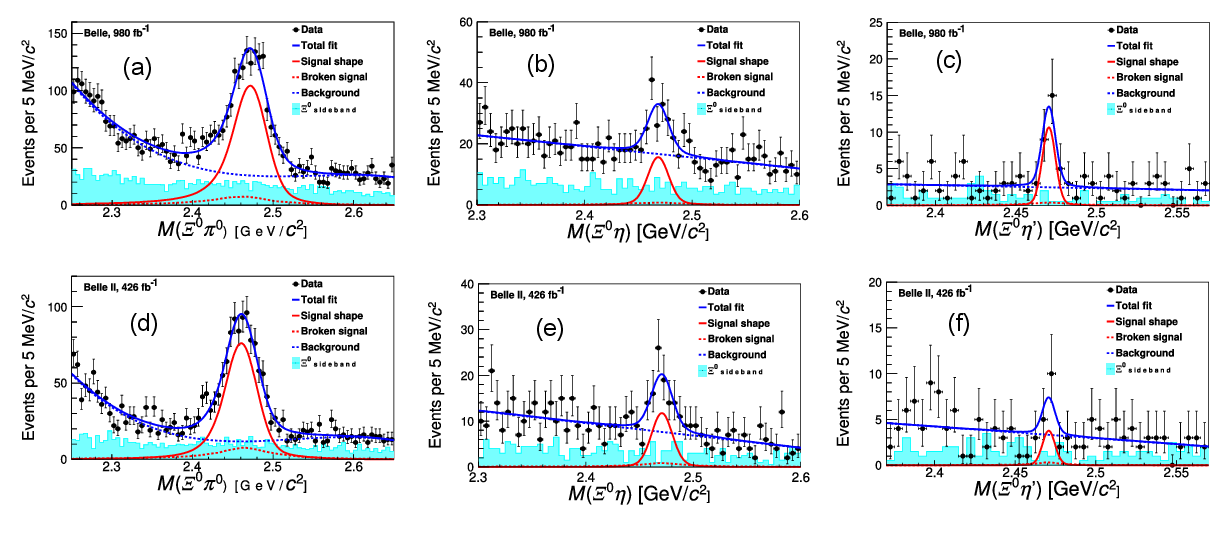} 
\caption{\label{fig:wide_last}
The invariant mass spectra of $\Xi_c^0$ candidates from (a, d) $\Xi_c^0 \to \Xi^0\pi^0$, (b, e) $\Xi_c^0 \to \Xi^0\eta$, and (c, f) $\Xi_c^0 \to \Xi^0\eta'$ decays reconstructed in Belle (top) and Belle II (bottom) data.}
\end{figure*}

The signal yields used for branching fraction measurements are extracted from fits to the invariant mass spectra of fully reconstructed $\Xi_c^0$ candidates (Fig.~\ref {fig:wide_last}). 
The measured values in the studied modes are 
\begin{subequations}
\label{eq:whole2}
\begin{eqnarray}
        \frac{\mathcal{B}(\Xi_c^0 \to \Xi^0 \pi^0)}{\mathcal{B}(\Xi_c^0 \to \Xi^- \pi^+)} = (0.48 \pm 0.02 \pm 0.03) \%, \\
      \frac{\mathcal{B}(\Xi_c^0 \to \Xi^0 \eta)}{\mathcal{B}(\Xi_c^0 \to \Xi^- \pi^+)}  = (0.11 \pm 0.01 \pm 0.01) \%,  \\
    \frac{\mathcal{B}(\Xi_c^0 \to \Xi^0 \eta')}{\mathcal{B}(\Xi_c^0 \to \Xi^- \pi^+)}  =
        (0.08 \pm 0.02 \pm 0.01) \%.
\end{eqnarray}
\end{subequations}

In the $\Xi^0_c \to \Xi^0 \pi^0$ decays, the asymmetry parameter $\alpha_c$ can be extracted from a fit to the $\Xi^0_c$ decay angular distribution, using the differential decay rate function:
\begin{eqnarray}
 \frac{dN}{d \cos\theta_{\Xi^0} } \varpropto 1+ \alpha_c \cdot \alpha_0 \cdot\cos\theta_{\Xi^0},
\end{eqnarray}
where $\alpha_0$ is the asymmetry parameter for the $\Xi^0 \to \Lambda\pi^0$ from cascade decay of $\Xi_c^0$ and $\theta_{\Xi^0}$ is the angle between the $\Lambda$ momentum vector and the  opposite direction to the $\Xi_c^0$ momentum vector in the $\Xi_c^0$ rest frame.

\begin{figure*}
\includegraphics[width = 0.7\linewidth]{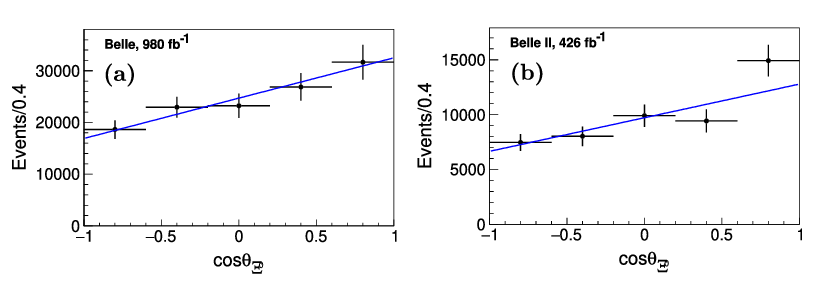} 
\caption{\label{fig:cosT}
Efficiency-corrected $\Xi_c^0$ signal yields in bins of $cos\theta_{\Xi^0}$ from the (a) Belle and (b) Belle II datasets. The lines show linear regression results.}
\end{figure*}

The asymmetry parameter $\alpha_c$ is obtained from a  fit to the $\Xi_c^0$ signal yield as a function of $\cos \theta_{\Xi^0}$ (Fig.~\ref{fig:cosT}).
Using the $\alpha_0$ value \cite{PDG}, the asymmetry parameter of $\Xi_c^0 \to \Xi^0 \pi^0$ is
\begin{equation}
    \alpha_c= -0.90\pm 0.15 \pm 0.23.
\end{equation}

Comparing the measured values of $\mathcal{B}(\Xi_c^0 \to \Xi^0 h^0)$ and asymmetry parameter $\alpha_c$ with the theoretical predictions~\cite{t6,t7},
we conclude that the $SU(3)_f$ model is the most consistent with experimental result.

\section{\label{sec:level1}Conclusion}

As can be seen from the presented analyses, charmed baryons remain an interesting object of study at the Belle and Belle II experiments. Several new decays of $\Xi_c^+$ and $\Xi_c^0$  were observed for the first time. The accuracy of $\mathcal{B}(\Lambda_c^+ \to pK^0_S\pi^0)$ was improved by a factor of five in comparison to the previous measurement. 
The asymmetry parameter $\alpha_c$ in the $\Xi_c^0 \to \Xi^0 \pi^0$ decays, which determines the decay dynamics, was measured for the first time. New Belle~II results allow discrimination of theoretical models.

\begin{acknowledgments}
The article was prepared within the framework of the project "Mirror Laboratories" HSE University.
\end{acknowledgments}

\nocite{*}


\end{document}